\documentclass[twocolumn,aps,showpacs,prl,superscriptaddress,floatfix]{revtex4}

\usepackage{graphicx}
\usepackage{amsmath,amssymb}
\usepackage{bm}

\def\mpi{Max-Planck-Institut f\"ur Kernphysik, Saupfercheckweg 1, D-69117 Heidelberg, Germany}
\def\tamu{Institute for Quantum Studies and Dept. of Physics,
Texas A\&M University, College Station, Texas 77843-4242}
\def\princeton{Princeton Institute for Materials Research, Princeton University, Princeton, NJ 08544-1009}

\begin{document}

%\title{Measurement of the separation between molecules beyond classical limit}
\title{Measurement of the separation between atoms beyond diffraction limit}
\author{Jun-Tao Chang}
\affiliation{\tamu}

\author{J\"org Evers}
\affiliation{\mpi}
\affiliation{\tamu}

\author{Marlan O. Scully}
\affiliation{\tamu}
\affiliation{\princeton}

\author{M. Suhail Zubairy}
\email{zubairy@physics.tamu.edu}
\affiliation{\tamu}

\begin{abstract}
Precision measurement of
small separations between two atoms or molecules has been of
interest since the early days of science. Here, we
discuss a scheme which yields spatial information on a system of
two identical atoms placed in a standing wave laser field. The
information is extracted from the collective resonance
fluorescence spectrum, relying entirely on
far-field imaging techniques. Both the interatomic separation and
the positions of the two particles can be measured with
fractional-wavelength precision over a wide range of distances
from about $\lambda/550$ to $\lambda/2$.
\end{abstract}

\pacs{42.50.Ct, 42.30.-d, 42.50.Fx}
% 42.50.Ct Quantum description of interaction of light and matter; related
% 42.30.-d Imaging and optical processing
% 42.50.Fx Cooperative phenomena in quantum optical systems

\maketitle

%%%%%%%%%%%%%%%%%%%%%%%%%%%%%%%%%%%%%%%%%%%%%%%%%%%%%%%%%%%%%%%%
% INTRODUCTION
%%%%%%%%%%%%%%%%%%%%%%%%%%%%%%%%%%%%%%%%%%%%%%%%%%%%%%%%%%%%%%%%
%%revised paragraph
The measurement of small distances is a fundamental problem since
the early days of science. It has become even more important due
to recent interest in nanoscopic and mesoscopic
phenomena~\cite{mini}. Starting from the invention of the optical
microscope around 400 years ago, today's optical microscopy
methodologies can basically be divided into lens-based and
lensless imaging. In general, far-field imaging is lens-based and
thus limited by criteria such as the Rayleigh diffraction limit
which states that the achievable resolution in the focus plane is
limited to half of the wavelength of illuminating light. Further
limitation arises from out-of-focus light, which affects the
resolution in the direction perpendicular to the focal plane.
Many methods have been suggested to break these
limits~\cite{near-field,sedm,
entanglement,interferometry,multiphoton,non-identical,hettich}.
Lens-based techniques include confocal, non-linear femtosecond or
stimulated emission depletion microscopy~\cite{sedm}. Also
non-classical features such as entanglement, quantum
interferometry or multi-photon processes can be used to enhance
resolution~\cite{entanglement,interferometry,multiphoton}.
A particularly promising development is lensless near-field optics,
which can achieve nanometer spatial resolution~\cite{near-field}. Roughly speaking,
the idea is to have light interactions close enough to the object
to avoid diffraction. This, however, typically restricts near-field
optics to objects on a surface. In 1995, Betzig proposed a method
to reach sub-wavelength resolution that is not limited to one spatial
dimension by assuming non-identical, individually addressable
objects~\cite{non-identical}.
Subsequently, this was realized in a landmark experiment of
Hettich et al.~\cite{hettich}. It combined near-field and
far-field fluorescence spectroscopy techniques, using the
fluorescence spectrum to label different molecules inside an
inhomogeneous external electrical field. They also
noticed dipole-dipole interactions between
adjacent objects~\cite{Dicke,ficekbook,spectrum} and used it to
correct the measurement result.
However, there is still great interest in achieving nanometer
distance measurements by using optical illuminating far-field
imaging only.

%%%%%%%%%%%%%%%%%%%%%%%%%%%%%%%%%%%%%%%%%%%%%%%%%%%%%%%%%%%%%%%%
\begin{figure}[b]
\includegraphics[width=8cm]{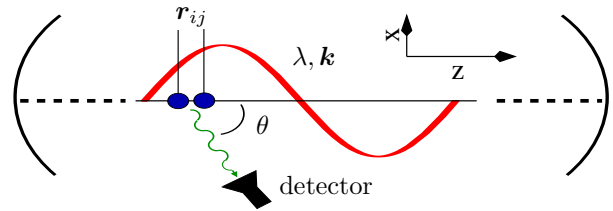}
\caption{\label{fig-system}(Color online) Two atoms in a standing wave field
 separated by a distance $|{\bf r}_{ij}|$ smaller than half of the
wavelength $\lambda$ of the driving field. The distance of the two
atoms is measured via the emitted resonance fluorescence.}
\end{figure}
%%%%%%%%%%%%%%%%%%%%%%%%%%%%%%%%%%%%%%%%%%%%%%%%%%%%%%%%%%%%%%%%
%

In this communication, we propose a scheme to measure the distance
between two adjacent two-level systems by driving them with a
standing wave laser field and measuring the far field resonance
fluorescence spectrum, which is motivated by the localization of
single atom inside a standing wave
field~\cite{interference,localization}. In particular, we focus on
distances smaller than the Rayleigh limit $\lambda/2$.
Our basic approach is that in a standing wave, the effective
driving field strength depends on the position of the particles.
Thus, each particle generates a sharp sideband peak in the
spectrum, where the peak position directly relates to the
subwavelength position of the particle. As long as the two
sideband peaks can be distinguished from each other, the position
of each particle can be recovered.
However, when the interatomic distance decreases, the two
particles can no longer be considered independent. Due to the
increasing dipole-dipole interaction between the two particles,
the fluorescence spectrum becomes complicated.
We find, however, that the dipole-dipole interaction energy can
directly be extracted from the fluorescence spectrum by adjusting
the parameters of the driving field.
Since the dipole-dipole interaction energy is distance dependent,
it yields the desired distance information.
We provide detailed measurement procedures and our estimates show
that the scheme is applicable to inter-particle distances in a
 very wide range from $\lambda/2$ to about $\lambda / 550$.

%%%%%%%%%%%%%%%%%%%%%%%%%%%%%%%%%%%%%%%%%%%%%%%%%%%%%%%%%%%%%%%%
% Setup
%%%%%%%%%%%%%%%%%%%%%%%%%%%%%%%%%%%%%%%%%%%%%%%%%%%%%%%%%%%%%%%%
Our model system consists of two identical two-level atoms located
at fixed points ${\bf r}_i=(x_i,y_i,z_i)^T$ ($i=1,2$) in a
resonant standing wave laser field (see Fig.~\ref{fig-system}).
The atomic transition frequency is $\omega_0$. The laser field has
frequency $\omega_L$, wavelength $\lambda$  and wave vector ${\bf
k}=k\hat{{\bf z}}$. We assume the two atoms to be arranged along
$\hat{\bf z}$. The driving field Rabi frequency of atom $i$ is
$\Omega_i$, with $\Omega_i = \Omega\,\sin (k\cdot z_{i})$. We
denote the raising (lowering) operator of the $i$th atom by
$S_i^+$ ($S_i^-$).
In the following, we assume the transition dipole moments of the
two atoms to be parallel and aligned perpendicular to the
$\hat{\bf z}$ direction. We also assume resonant driving, $\Delta
= \omega_L-\omega_0 = 0$.

If the two atoms are far apart (the distance between atom $i$ and
$j$ $|z_{ij}|=|{z}_i -{z}_j|\gg \lambda$), then they are
independent, and the total Master equation is given by the sum of
the two single-particle Master equations~\cite{ficekbook}. 
%%%%%%%%%%
If the two atoms come close, they dipole-dipole
interact, causing a collective system dynamics. This gives rise
to a complex energy shift due to a virtual photon exchange
between the two atoms. 
The imaginary part of this shift corresponds to an incoherent coupling,
whereas the real part shifts the energy of the collective states
of the system.
%%%%%%%%
The full
collective Master equation is given by~\cite{Dicke,ficekbook}
\begin{align}
\frac{\partial \rho}{\partial t} =& \frac{1}{i\hbar} [H, \rho]
- \sum_{i,j=1}^{2} \gamma_{ij} \left ([S_i^+,S_j^-\rho] -
[S_j^-,\rho S_i^+] \right )\,. \label{master}
\end{align}
The coherent evolution is governed by $H=H_0 + H_{dd} + H_L$.
The free energy $H_0 = (\hbar/2)\omega_0 \sum_{i=1}^{2}( S_i^+S_i^- - S_i^-S_i^+ )$
of the two atoms and the interaction with the driving laser field
$H_{L} = (\hbar /2) \sum_{i=1}^{2} \left (\Omega_i S_i^+
e^{-i \omega_L t} + \textrm{H.c.} \right )$ are the same as for two independent
atoms. The coherent energy shift of the collective states arises from the
dipole-dipole interaction $H_{dd} =  \hbar\Omega_{12} (S_1^+S_2^- +
\textrm{H.c.})$, which involves couplings of both atoms.
For the considered geometry, the dipole-dipole interaction
$\Omega_{12}$ is given by
\begin{align}
\Omega_{12}=&\frac{3}{2}\gamma \left \{
-\frac{\cos(kz_{ij})}{(kz_{ij})} +
\frac{\sin(kz_{ij})}{(kz_{ij})^2}+\frac{\cos(kz_{ij})}{(kz_{ij})^3}
\right \}\,. \label{omega12}
\end{align}
%
%Here, $z_{ij} = |z_i - z_{j}|$ is the
%distance between atom $i$ and $j$.
The incoherent evolution first entails the independent spontaneous emission
of the two atoms $\gamma_{ii}$ ($i=1,2$), as found for uncoupled atoms.
Terms with $\gamma_{ij}$ ($i\neq j$) are the incoherent dipole-dipole
couplings, where
\begin{align}
\gamma_{ij}=&\frac{3}{2}\gamma \left \{
\frac{\sin(kz_{ij})}{(kz_{ij})} +
\frac{\cos(kz_{ij})}{(kz_{ij})^2}-\frac{\sin(kz_{ij})}{(kz_{ij})^3}
\right \}\,. \label{gammaij}
\end{align}
For large distances, ($k z_{ij} \gg 1$), we find
$\Omega_{12}\approx 0$ and $\gamma_{ij}\approx\gamma \delta_{ij}$,
where $\delta_{ij}$ is the Kronecker Delta symbol.
Thus we recover the case of two independent atoms, as expected.
For small distances ($k z_{ij} \ll 1$), one finds maximum
incoherent cross-coupling, and $\Omega_{12}$ approaches
the static dipole-dipole interaction,
\begin{align}
\Omega_{12}\approx&3\gamma / [2 (kz_{ij})^3]
%[1-3\cos^2\alpha]
\,, \qquad \gamma_{ij}\approx \gamma\,. \label{param-approx}
\end{align}
%
%
%%%%%%%%%%%%%%%%%%%%%%%%%%%%%%%%%%%%%%%%%%%%%%%%%%%%%%%%%%%
\begin{figure}[t]
\includegraphics[width=8cm]{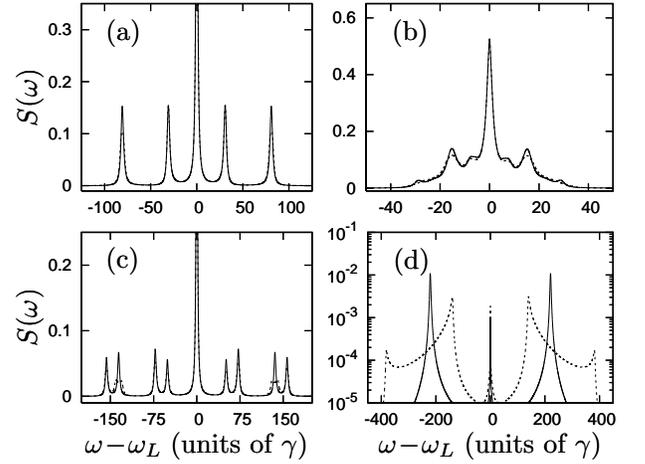}
\caption{\label{sample-spectra}Sample spectra for $\Delta=0$,
 $\theta=\pi/2$, $z_{1}=0.05\lambda$. Fixed distance $z_{12}$
 (solid lines) and harmonic oscillation around $z_{12}$ (dashed lines).
(a) Large separation case: $z_{12}=0.3\lambda, \Omega =
100\gamma$.
(b) Intermediate separation, weak driving field:
 $z_{12}=0.08\lambda, \Omega = 20\gamma$
(c) As (b), but strong driving field:
   $z_{12}=0.08\lambda, \Omega = 200\gamma$
(d) Small separation: $z_{12}=0.03\lambda, \Omega =
20\gamma$.}
\end{figure}
%%%%%%%%%%%%%%%%%%%%%%%%%%%%%%%%%%%%%%%%%%%%%%%%%%%%%%%%%%%
%
%%%%%%%%%%%%%%%%%%%%%%%%%%%%%%%%%%%%%%%%%%%%%%%%%%%%%%%%%%%
%% fluorescence spectrum
%%%%%%%%%%%%%%%%%%%%%%%%%%%%%%%%%%%%%%%%%%%%%%%%%%%%%%%%%%%
Our strategy is to identify the distance of the two atoms via
the emitted resonance fluorescence. 
We define
$\hat{\bf R}$ as the unit
vector in observation direction, and the observation angle
$\theta$ as $\theta = \arccos (\hat {\bf R}\cdot {\bf r}_{12}/r_{12})$.
The total two-atom steady state
resonance fluorescence spectrum
$S(\omega)$  up to a geometrical factor
is given by~\cite{ficekbook}
\begin{align*}
S(\omega) = {\rm Re}&\,\int_0^\infty d\tau
e^{i(\omega-\omega_L)\tau}
\sum_{i,j=1}^{2}
\langle S_i^+(0)S_j^-(\tau)\rangle_s \: e^{ik\hat{\bf R}\cdot{\bf r}_{ij}}\,,
\end{align*}
where the subindex $s$ denotes the steady state. 
In general, this resonance fluorescence spectrum
is rather complicated~\cite{spectrum}. The
spectrum, however, simplifies considerably in limiting cases,
where either the driving field Rabi frequency or the dipole-dipole
interaction dominates the dynamics. This will be
exploited in the following, where we present in detail a
measurement procedure, which allows us to extract the distance
between the two atoms and their positions relative to nodes of the
standing wave field, both with fractional-wavelength precision.
%
%
%%%%%%%%%%%%%%%%%%%%%%%%%%%%%%%%%%%%%%%%%%%%%%%%%%%%%%%%%%%
%% measurement scheme
%%%%%%%%%%%%%%%%%%%%%%%%%%%%%%%%%%%%%%%%%%%%%%%%%%%%%%%%%%%
The first step in the measurement sequence is to apply a
standing wave laser field to the two atoms, which at an anti-node of the
standing wave corresponds to a Rabi frequency $\Omega$ of a few $\gamma$.
Depending on the relative separation of the atoms, different spectra
can be observed.

\begin{figure}[t!]
\includegraphics[width=8cm]{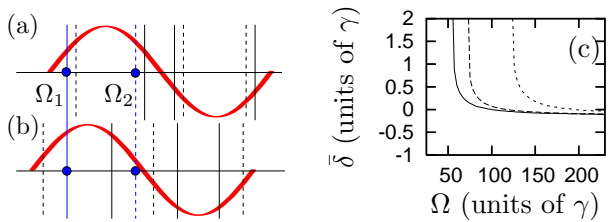}
\caption{\label{intermediate-peakpos} (Color online) (a,b) Obtaining the position of the two atoms via a
phase shift of the standing wave field. Solid (dashed) lines show
possible atom positions for given $\Omega_1$ ($\Omega_2$). (a)
Before, (b) after the phase shift. The only coinciding potential
positions in (a) and (b) give the true atomic positions. (c) Deviation $\bar{\delta}
= \sigma_{\rm p} - 2\Omega_{12}$ of the doublet splitting
$\sigma_{\rm p}$ from $2\Omega_{12}$ for the strong field,
intermediate distance case. $z_{12} = 0.08\lambda$,
$\theta=\pi/2$, and $\Delta=0$. The positions of the atoms are
$z_{1}=0.05\lambda$ (solid), $0.1\lambda$ (dashed), $0.15 \lambda
$ (dotted). }
\end{figure}

%%%%%%%%%%%%%%%%%%%
%% large-distance case
%%%%%%%%%%%%%%%%%%%
If the two atoms are well-separated (about $\lambda/10 \lesssim
z_{12}\lesssim \lambda/2$), then the dipole-dipole interaction
is negligible. In this case spectra as shown in
Fig.~\ref{sample-spectra}(a) are obtained. The two sideband
structures can be interpreted as arising from the AC-Stark
splitting due to $\Omega_1$ and $\Omega_2$. Thus the sideband peak
positions $\nu^{p}_{1}$ and $\nu^{p}_{2}$ can directly be related
to $\Omega_1$ and $\Omega_2$ and therefore to the position of the
two atoms relative to the standing wave field nodes. Consequently,
we can obtain the distance $z_{12}$.
Within half a
wavelength, however, in general two interatomic distances are
possible for measured values of $\Omega_1$ and $\Omega_2$ (see
Fig.~\ref{intermediate-peakpos}(a,b))~\cite{phase}. An
identification of the actual atomic positions is possible by
changing the standing wave phase slightly, i.e., shifting
the positions of the (anti-) nodes. As shown in
Fig.~\ref{intermediate-peakpos}(a,b), a combination of the
possible positions for two different standing wave phases
yields the actual separation.
%It may also be possible
%to use alternative schemes to restrict possible positions for the
%atoms, e.g. using phase-dependent schemes as discussed
%in~\cite{phase} for a single atom.
Note that this complication is
not present for nearby atoms, where the non-vanishing
dipole-dipole energy allows to determine the distance directly.
In Fig.~\ref{sample-spectra}(a), the distance of the
two particles is $z_{12}=0.3\lambda$.
From the spectrum accessible in experiments, the distance
$z^{exp}_{12}=( 0.300\pm 0.02)\lambda$ is obtained,
if we allow for a total measurement uncertainty of about $10\%$.
Thus the actual and the measured distances match, and the
$10\%$ uncertainty of the distance measurement corresponds
to about $\lambda /50$.

%%%%%%%%%%%%%%%%%%%%%%%%%%%%%%%%%%%%%%%%%%%%%%%%%%%%%%%%%%%%%%%%%%%%%
%%%%%%%%%%%%%%%%%%%%%%%%%%%%%%%%%%%%%%%%%%%%%%%%%

%%%%%%%%%%%%%%%%%%%
%% indecisive initial measurement
%%%%%%%%%%%%%%%%%%%
If the distance between the two atoms is intermediate (about
$\lambda/30 \lesssim z_{12}\lesssim \lambda/10$), then the initial
weak field measurement in general yields a more complicated
spectrum, see Fig.~\ref{sample-spectra}(b). The
reason is that then the dipole-dipole coupling and
the driving field strength are comparable, and the two atoms
are not independent.
In such a case, a quantitative interpretation of the spectrum is
difficult. However, increasing the Rabi frequency $\Omega$ leads
to a spectrum as shown in Fig.~\ref{sample-spectra}(c). The
spectrum consists of a central peak, two inner sideband doublets,
and two outer sideband doublets, each symmetrically placed around
the driving field frequency $\omega_L$. The center positions of
the inner and outer sideband doublets corresponds to the Rabi
frequencies $\Omega_1$ and $\Omega_2$. The sideband structures are
split into doublets due to the dipole-dipole coupling of the two
atoms. For large $\Omega$, the splitting approaches twice the
energy $\Omega_{12}$, as shown in
Fig.~\ref{intermediate-peakpos}(c). Thus the strong-field sideband
doublet splitting directly yields $\Omega_{12}$ and then the
distance of the two atoms, via Eqs.~(\ref{omega12}).% or
%(\ref{param-approx}).
%
For example, in
Fig.~\ref{sample-spectra}(c), the actual distance is
$z_{12}=0.08\lambda$. From the spectrum, a measurement would obtain
$\Omega_{12}=(10.54\pm 1.05)\gamma$, where again we have allowed
for an uncertainty of about $10\%$.
From Eq.~(\ref{omega12}), this
yields a measured distance of $z_{12}=(0.0801\pm0.0027)\lambda$,
in good agreement with the actual value.
%
%In the strong field
%limit, the mean frequency of the two peaks of each sideband
%structure corresponds to the Rabi splitting $\Omega_1$ or
%$\Omega_2$, respectively, such that (from a comparison with
%$\Omega$) the positions of the individual atoms relative to
%standing wave field nodes can be obtained.
On the other hand, comparing the center center positions of the
inner and outer sideband doublets with $\Omega$, the positions of
the individual atoms relative to standing wave field nodes can be
obtained. For the setup in Fig.~\ref{sample-spectra}(c), we have
$z_{1} = 0.05\lambda$, $\Omega_1 = 61.80\gamma$,
$\Omega_2=145.79\gamma$. From the spectrum, using the above
procedure, we obtain $\Omega_1 = (61.58\pm 6.16)\gamma$,
$\Omega_2=(146.22\pm 14.62)\gamma$, assuming a relative
uncertainty of $10\%$. From $z_{1}= \lambda /2\pi
\arcsin(\Omega_1/\Omega)$, this would yield a measurement result
of $z_{1} = (0.050\pm0.005)\lambda$, in good agreement with the
actual position of the atoms.

In the above two regimes, the situation slightly complicates if
both atoms are located near-symmetrically around a node or an
anti-node. In this case, $\Omega_1 \approx \Omega_2$, such that
the two sideband peaks (or doublets) overlap. One way to resolve
this is to adequately change the standing wave field phase. By
this, the symmetry can be lifted to give $\Omega_1 \neq \Omega_2$.
Then the above procedure can be applied to yield the separation
and positions.

\begin{figure}[t!]
\includegraphics[width=8cm]{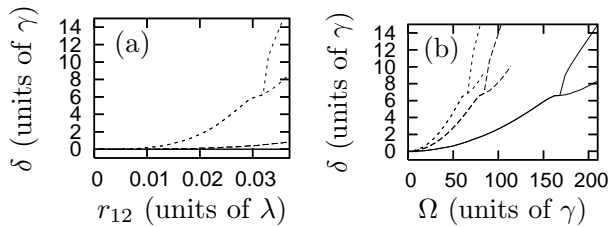}
\caption{\label{peakpos-small-distance}Deviation $\delta =
\nu_{\rm p} - \Omega_{12}$ of the peak position $\nu_{\rm p}$ from
$\Omega_{12}$ for closely-spaced atoms. $\Delta=0$,
$\theta=\pi/2$, and (a) Against the atomic separation.
$z_{1}=0.05\lambda$, $\Omega = 3\gamma$ (solid), $20\gamma$
(dashed), $80\gamma$ (dotted). (b) Against the driving
field Rabi frequency. $z_{12} = 0.02\lambda$, $z_{1}=0.05
\lambda$ (solid), $0.125 \lambda$ (dashed), $0.2 \lambda$ (dotted).
Branches indicate splittings into two peaks.}
\end{figure}

%%%%%%%%%%%%%%%%%%%
%% small-distance case
%%%%%%%%%%%%%%%%%%%
If the two atoms are very close to each other (distance $\lesssim
\lambda / 30$), then the spectrum is dominated by the
dipole-dipole interaction energy $\Omega_{12}$, which gives rise
to sideband structure at each side of the fluorescence spectrum
close to $\omega_L \pm \Omega_{12}$, and only weakly depends on
the driving field. A typical spectrum for this parameter range is
shown in Fig.~\ref{sample-spectra}(d). As long as $\Omega_1$,
$\Omega_2$, $\gamma \ll \Omega_{12}$ is satisfied, the sideband
structures only have a small residual dependence on the Rabi
frequency. Thus, the sideband peak position $\nu_{\rm p}$ can
directly be identified with $\Omega_{12}$.
Fig.~\ref{peakpos-small-distance}(a) shows the deviation of the
sideband peak positions from $\Omega_{12}$ versus the atomic
separation distance for different Rabi frequencies $\Omega$. Note
that the effective Rabi frequencies $\Omega_1$, $\Omega_2$ also
depend on the position of the first atom within the wavelength,
with maximum values $\Omega_1, \Omega_2 \approx \Omega$ close to
the anti-nodes. It can be seen that for weak $\Omega_1, \Omega_2$,
the experimentally accessible sideband peak position and
$\Omega_{12}$ coincide very well. With increasing Rabi frequency,
the deviation increases, until the driving field induces a
splitting of the sideband peaks, indicated by the branching point
in Fig.~\ref{peakpos-small-distance}(a). If the initial spectrum
of the first measurement has insufficient signal-to-noise ratio,
then the fluorescence intensity can be enhanced by increasing the
driving field intensity. Note that due to the dependence of
$\Omega_1$, $\Omega_2$ on the position of the two atoms, different
positions of the two atoms may require different laser field
intensities. It is also possible to extrapolate the result of
several measurements to the driving field-free limit to increase
the measurement accuracy.
%
%%%%%%%%%%%%%%%%%%%%%%%%%%%%%%%%%%%%%%%%%%%%%%%%%%%%%%%%%%%%%%%%%%%%%
%%%%%%%%%%%%%%%%%%%%%%%%%%%%%%%%%%%%%%%%%%%%%%%%%%%%%%%%%%%%%%%%%%%%%
%
Via Eqs.~(\ref{omega12}) or (\ref{param-approx}), the measured
$\Omega_{12}$ can easily be used to obtain the interatomic
separation. The separation is measured with increasing accuracy in
the region of large slope of $\Omega_{12}$ vs $z_{12}$. For
maximum accuracy, Eq.~(\ref{omega12}) should be numerically solved
for the separation. Here, we discuss the small separation limit
Eq.~(\ref{param-approx}), and allow for a small uncertainty in
$\Omega_{12}$ ($\Omega_{12} \to \Omega_{12} + \delta
\Omega_{12}$). We obtain
%
%\begin{align}
%z_{ij} = \left ( \frac{3\gamma}{2k^3\Omega_{12}} \right)^{1/3}
%\left (1 - \frac{\delta \Omega_{12}}{3\Omega_{12}} \right )
%\end{align}
%
%
$z_{ij} = [3\gamma/(2k^3\Omega_{12})]^{1/3}
\cdot[1 - \delta \Omega_{12}/(3\Omega_{12})]$
as the distance $z_{ij}$ between the two atoms. Thus, the relative
uncertainty of the final result is about 1/3 of the relative
uncertainty of the measured $\Omega_{12}$. Consider, for example,
the case shown in Fig.~\ref{sample-spectra}(d). The actual
distance is $z_{12}=0.03\lambda$. The measured dipole-dipole
energy is $\Omega_{12}=(220.5\pm 22)\gamma$, again with a relative
measurement uncertainty of about $10\%$. From Eq.~(\ref{omega12}),
the distance then evaluates to $z_{12}=(0.030\pm 0.001)\lambda$.
Thus in this case, the uncertainty of the distance measurement
would be about $\lambda/1000$, i.e., less than $4\%$ of the actual
distance.

Once the distance $z_{12}$ is known, the positions of the two
atoms relative to nodes of the standing wave field can be
obtained. For this, we note from
 Fig.~\ref{peakpos-small-distance}(b) that---for otherwise fixed
parameters---the position of the branching point depends on the
Rabi frequencies $\Omega_1$, and thus on the position $z_{1}$. If
in the experiment we increase $\Omega$ up to the branching point,
then the position of the atom pair relative to the field nodes can
be deduced. Accurate analytic expressions for the position of the
branching point, however, are involved, as the general expression
of the fluorescence spectrum is
complicated~\cite{spectrum}. Thus a numerical fit as shown in
Fig.~\ref{peakpos-small-distance}(b) should be used to evaluate
 $z_{1}$. Finally, since for
small distances the spectrum is almost independent to the
driving field, the distance information can also be obtained using
a travelling-wave field, which may be more convenient in practice.

The precise positioning of the atoms is
limited by thermal or quantum position
uncertainties~\cite{position}. We have simulated this effect
by assuming a motional ground state harmonic oscillation with 
amplitude $\delta z_{12}=0.005 \lambda$
(corresponding to a Lamb-Dicke parameter $\eta\approx 0.016$)
around the mean distance $z_{12}$. Results averaged over this motion
are shown with dashed lines in Fig.~\ref{sample-spectra}.
For $\delta z_{12} \ll z_{12}$, the motion is negligible. With increasing
ratio $\delta z_{12}/z_{12}$, the spectral peaks split up. In 
Fig.~\ref{sample-spectra}(d), two peaks emerge at the classical
turning points of the distance oscillation. From these, the mean distance
can again be obtained.
%
%%%%%%%%%%%%%%%%%%%
%% limitations
%%%%%%%%%%%%%%%%%%%
The possible separation measurement range is limited, as the
dipole-dipole coupling $\Omega_{12}$ increases with decreasing
separation as $z_{12}^{-3}$. For our model to remain valid,
however, $\Omega_{12}\ll \omega_0$ should be fulfilled. From
Eq.~(\ref{param-approx}), for $\gamma\sim 10^7$ Hz, $\Omega_{12}
\leq 10^{13}$ Hz, we estimate $z_{12}\geq \lambda / 550$ as the
theoretical resolution limit. This limitation only applies to the
distance of the two atoms itself; the distance uncertainty in
principle can be well below $\lambda /550$. Note 
that these considerations neglect experimental uncertainties,
and are subject to imperfections e.g. in the measurement of laser
field parameters or the alignment of dipole moments or
laser fields.

%%%%%%%%%%%%%%%%%%%%%%%%%%%%%%%%%%%%%%%%%%%%%%%%%%%%%%%%%
% Summary
%%%%%%%%%%%%%%%%%%%%%%%%%%%%%%%%%%%%%%%%%%%%%%%%%%%%%%%%%
In summary, we have discussed a microscopy scheme entirely
based on optical far-field techniques.
It allows to measure the
separation between and the position of two nearby atoms in a
standing wave laser field with fractional-wavelength precision
over the full range of distances from about $\lambda/550$ up to
the Rayleigh limit $\lambda/2$.

%%%%%%%%%%%%%%%%%%%%%%%%%%%%%%%%%%%%%%%%%%%%%%%%%%%%%%%%%
% Acknowledgements
%%%%%%%%%%%%%%%%%%%%%%%%%%%%%%%%%%%%%%%%%%%%%%%%%%%%%%%%%
\section*{Acknowledgements}
JE thanks for hospitality during his stay at Texas A\&M University.
This research is supported by the Air Force Office of Scientific Research,
DARPA-QuIST, Office of Naval Research, and the TAMU Telecommunication and
Informatics Task Force (TITF) initiative.

\end{document}